\documentclass[12pt,epsfig]{article}
\input epsf
\begin{document}
\title{$\pi\pi$ scattering $S$ wave from the data on the reaction
$\pi^-p\to\pi^0\pi^0n$\\[0.7cm]}
\author{N.N. Achasov \thanks{Email: achasov@math.nsc.ru}\,\,
and G.N. Shestakov \thanks{Email: shestako@math.nsc.ru}\\[0.5cm]
{\it Laboratory of Theoretical Physics,}\\ {\it S.L. Sobolev
Institute for Mathematics,}\\ {\it 630090, Novosibirsk, Russia}}
\date{}
\maketitle
\begin{abstract}
The results of the recent experiments on the reaction
$\pi^-p\to\pi^0\pi^0n$ performed at KEK, BNL, IHEP, and CERN are
analyzed in detail. For the $I=0$ $\pi\pi$ $S$ wave phase shift
$\delta^0_0$ and inelasticity $\eta^0_0$ a new set of data is
obtained. Difficulties emerging when using the physical solutions
for the $\pi^0\pi^0$ $S$ and $D$ wave amplitudes extracted with
the partial wave analyses are discussed. Attention is drawn to the
fact that, for the $\pi^0\pi^0$ invariant mass, $m$, above 1 GeV,
the other solutions, in principle, are found to be more preferred.
For clarifying the situation and further studying the $f_0(980)$
resonance thorough experimental investigations of the reaction
$\pi^-p\to\pi^0\pi^0n$ in the $m$ region near the $K\bar K$
threshold are required.
\end{abstract}
\newpage
\begin{center}{\bf I. INTRODUCTION}\end{center}

The reactions $\pi N\to\pi\pi N$ so far are a major source of
information on the processes $\pi\pi\to\pi\pi$. At high energies
and small values of the momentum-transfer-squared from the
incident $\pi$ to the outgoing $\pi\pi$ system, $0<-t<0.2$
GeV$^2$, the reactions $\pi N \to\pi\pi N$ are dominated by the
one-pion exchange (OPE) mechanism. In treating the data on these
reactions the partial wave analysis method is used. As a rule, a
few possible solutions for the partial wave amplitudes of the
final $\pi\pi$ system  are obtained. In some cases, the preferred
solution is selected from the additional physical arguments.
Generally, to obtain the reliable and unambiguous results in a
wide region of $m$, high statistics, polarized targets, and
precise measurements of the $\pi N\to\pi\pi N$ reaction cross
section at different energies are needed. The detailed reviews and
comprehensive discussions of the experimental results on the
reactions $\pi N\to\pi\pi N$ and $\pi\pi$ scattering in the region
$2m_\pi<m<2$ GeV available by the early 1999 have been presented
in Refs. [1,2].

In this work we analyze the recent data on the intensities and
relative phase of the $S$ and $D$ partial waves of the
$\pi^0\pi^0$ system produced in the reaction
$\pi^-p\to\pi^0\pi^0n$. The data have been obtained in the
experiments with incident $\pi^-$ energies of 8.9, 18.3, 38, and
100 GeV performed at KEK [3], BNL [4], IHEP [5], and CERN [6],
respectively. Our main goal is to obtain information on the
$\pi\pi$ $S$ wave phase shift $\delta^0_0$ and inelasticity
$\eta^0_0\,$ in the channel with isospin $I=0$ that would be
complementary to the previous ``canonical"\ data extracted from
the 17.2 GeV experiments on the reactions $\pi^-p\to\pi^+\pi^-n$
[7-12]. We especially emphasize a strong likeness of the physical
solutions selected in all four experiments on $\pi^0\pi^0$
production and also the common difficulties that emerge when
interpreting these solutions and in their comparison with the
$\pi^+\pi^-$ data. It turns out, in particular, that some of the
physical solutions found lead to considerable violations of the
unitarity condition for the $\pi\pi$ scattering amplitude in
question. In addition, we conclude that the data on $\pi^0\pi^0$
production are indicative of a noticeably smaller value of the
$f_2(1270)\to\pi\pi$ decay branching ratio in comparison with the
Particle Data Group (PDG) data [13]. In connection with a
considerable interest in the light scalar meson sector (see for
reviews Refs. [1,2,13-15]), we suggest to perform especially
careful measurements of the reaction $\pi^-p\to\pi^0\pi^0n$ in the
$m$ region from 0.9 to 1.1 GeV, i.e., near $K\bar K$ threshold.
This would allow the $f_0(980)$ coupling constant to the $K\bar K$
channel to be determined more reliably and also to resolve the
long-standing question [16] of a possible ambiguity in the
behavior of the phase shift $\delta^0_0$ above the $K\bar K$
threshold.

The paper is organized as follows. In Sec. II, the KEK results [3]
are analyzed. In Ref. [3] the data on the phase shift $\delta^0_0$
have been obtained in the $m$ region from 0.36 to 1 GeV. The
$\delta^0_0$ values found by us in other way in the interval
$0.68\leq m\leq1$ GeV agree with the KEK data [3] within
experimental uncertainties. We also present new results for
$\delta^0_0$ and $\eta^0_0$ in the region $1<m<1.64$ GeV. In Sec.
III, the extrapolation of the $S$ and $D$ wave mass distributions
obtained in the BNL experiment [4] from the physical region of the
reaction $\pi^-p\to\pi^0\pi^0n$ to the pion pole ($t=m^2_\pi$) is
performed. Considering the different solutions found in Ref. [4]
for these distributions we obtain a few sets of the values of
$\delta^0_0$ and $\eta^0_0$ in the $m$ region from 0.32 to 1.52
GeV. The GAMS results on the reaction $\pi^-p\to\pi^0\pi^0n$ [5,6]
are discussed in Sec. IV. Here we also summarize all difficulties
encountered while analyzing the $\pi^0\pi^0$ data [3-6]. In Sec.
V, we formulate briefly a few concrete suggestions for further
studying the reaction $\pi^-p\to\pi^0\pi^0n$ which, as one can
hope, will be used to clarify the experimental situation.

\begin{center}{\bf II. ANALYSIS OF THE KEK DATA}\end{center}

In the KEK experiment [3], the data on the intensities and the
relative phase of the $S$ and $D$ partial waves for the reaction
$\pi^+\pi^-\to\pi^0\pi^0$ have been obtained in the $m$ interval
from 0.36 to 1.64 GeV. They have been extracted from the
$\pi^-p\to\pi^0\pi^0n$ data by using the linear Chew-Low
extrapolation and partial wave analysis. Because the absolute
$\pi^0\pi^0$ production cross section has not been determined in
the experiment, the $S$ and $D$ wave intensities, $|A_S|^2$ and
$|A_D|^2$, have initially been presented in arbitrary (identical)
units [3]. Any alternative solution for $|A_S|^2$ and $|A_D|^2$
has not been discussed in Ref. [3]. The $S$ and $D$ wave
intensities are related to the phase shifts $\delta^I_0$ and
$\delta^I_2$ and inelasticities $\eta^I_0$ and $\eta^I_2$ in a
conventional way: $|A_S|^2\sim|a^0_0-a^2_0|^2$, where
$a^I_0=(\eta^I_0\exp(2i\delta^I_0)-1)/2i$, and
$|A_D|^2\sim|a^0_2-a^2_2|^2$, where
$a^I_2=(\eta^I_2\exp(2i\delta^I_2)-1)/2i$. To find $\delta^0_0$
below the $K\bar K$ threshold, it was assumed [3] that in this
region $\eta^0_0=\eta^2_0=1$, and  consequently
$|A_S|^2\sim\sin^2(\delta^0_0-\delta^2_0)$. As is well known from
a large number of previous experiments, the phase shift
$\delta^0_0$ smoothly goes through 90$^\circ$ in the region
$0.7<m<0.9$ GeV and the phase shift $\delta^2_0$ is negative,
smooth, and small (see, for example, [2,7,8,17]). Therefore, to
extract the phase shift difference $\delta^0_0-\delta^2_0$ from
the unnormalized data, the following normalization condition was
accepted in Ref. [3]: the maximum value of $|A_S|^2$ is equal to
1. The KEK data for the $S$ and $D$ partial wave intensities
normalized in this way ara shown in Fig. 1, together with the data
on the relative phase $\delta=\phi_S-\phi_D$ between the
amplitudes $A_S=|A_S|\exp{(i\phi_S)}$ and
$A_D=|A_D|\exp{(i\phi_D)}$. The values of the $I=2$ $\pi\pi$ $S$
wave phase shift $\delta^2_0$ used in Ref. [3] were given by the
parametrization $\delta^2_0=-0.87q$ [with
$q=(m^2/4-m^2_\pi)^{1/2}$ in GeV and $\delta^2_0$ in radians], and
in such a way the data for $\delta^0_0$ were obtained in the $m$
region from 0.36 to 1 GeV. In the following, for $\delta^2_0$ we
shall use the fit to the data from Refs. [17,18] which is shown in
Fig. 2. \footnote{This fit was obtained with the parametrization
$\delta^2_0=-aq/(1+bm^2+cm^4+dm^6)$, where $\,a,\ b,\ c\ $, and
$\,d$ are fitted parameters. In addition, in lacking the reliable
data on the deviation of $\eta^2_0$ from 1, we set $\eta^2_0=1$
for all considered values of $m$. When more or less detailed data
on $\eta^2_0$ are available, it will be interesting to take into
account possible inelastic effects. It is not unreasonable to
think that such effects may appear in the $I=2$ $\pi\pi$ channel
only above the nominal $\rho\rho$ threshold (1.54 GeV), but not
above the $K\bar K$ threshold as in the $I=0$ $\pi\pi$ channel.}
Using the fit and the data for $|A_S|^2$ shown in Fig. 1a we have
also determined the values of $\delta^0_0$ for $m<1$ GeV. They are
plotted in Fig. 1d. The resulting values are in excellent
agreement with those obtained in Ref. [3].

We now determine $\delta^0_0$ and $\eta^0_0$ simultaneously by
using the available data on the relative phase $\delta$ and the
intensity $|A_S|^2$ (see Figs. 1c and 1d) in the $m$ region from
0.68 to 1.64 GeV. In order to estimate the phase $\phi_D$ we
neglect the tiny amplitude $a^2_2$ [17,18] (which is quite
reasonable because the experimental errors of $|A_D|^2$, as is
seen from Fig. 1b, are not too small) and assume that the $D$ wave
amplitude is dominated by the $f_2(1270)$ resonance contribution
and can be written in the form
\begin{equation}A_D=\frac{m_{f_2}B_{f_2\pi\pi}\,\Gamma}{m^2_{f_2}-
m^2-im_{f_2}\Gamma}\,, \end{equation} were
$\Gamma=(m_{f_2}/m)\Gamma_{f_2}(q/q_{f_2})^5D(q_{f_2}R_{f_2})/D(qR_{f_2})$,
$D(x)=9+3x^2+x^4$, $q_{f_2}=(m^2_{f_2}/4-m^2_\pi)^{1/2}$,
$R_{f_2}$ is the interaction radius, and  $m_{f_2},\
\Gamma_{f_2}$, and $B_{f_2\pi\pi}$ are the mass, width, and
$\pi\pi$ decay branching ratio of the $f_2(1270)$. The fitted
curve on Fig. 1b corresponds to the following values of the
$f_2(1270)$ resonance parameters:
\begin{eqnarray}m_{f_2}=1.283\pm0.008\ \mbox{GeV}\,,& \qquad & \Gamma_{f_2}
=0.170\pm0.014\ \mbox{GeV}\,,\nonumber\\ R_{f_2}=3.59\pm0.71\
\mbox{GeV}^{-1}\,,& \qquad & B_{f_2\pi\pi}=0.760\pm0.034\,.
\end{eqnarray} Thus, the phase $\phi_D$ is defined by that of the Breit-Wigner
amplitude (1). To express the parameters $\delta^0_0$ and
$\eta^0_0$ in terms of the known values $\delta$, $\,|A_S|^2$,
$\,\delta^2_0\,$, and $\,\phi_D$, it is convenient to represent
the amplitude $A_S$ in the form (see footnote 1)
\begin{equation}A_S=e^{2i\delta^2_0}\left(\frac{\eta^0_0\,
e^{2i(\delta^0_0-\delta^2_0)}-1}{2i}\right)=e^{2i\delta^2_0}\widetilde{A}_S=
e^{i(2\delta^2_0+\phi)}|\widetilde{A}_S|\,,
\end{equation} where $\phi$ is the phase of the amplitude
$\widetilde{A}_S$. The distinctive feature of the amplitude
$\widetilde{A}_S$ is that in its Argand diagram the relations
between $\delta^0_0-\delta^2_0$, $\,\phi$, $\,\eta^0_0$, and
$|\widetilde{A}_S|$ formally look like the relations between the
corresponding parameters of any unitary partial wave amplitude
with definite isospin $I$; for example, the phase $\phi$ is
confined within the range from $0^\circ$ to $180^\circ$ because
Im($\widetilde{A}_S)>0$. Thus, we have
\begin{eqnarray}
\phi=\delta-2\delta^2_0+\phi_D\,,\ \ \ \ \  & \qquad &
\eta^0_0=\sqrt{1-4|A_S|\sin\phi+4|A_S|^2}\,,
\end{eqnarray} \begin{eqnarray}
\sin2(\delta^0_0-\delta^2_0)=\frac{2|A_S|\cos\phi}{\eta^0_0}\,, &
\qquad &
\cos2(\delta^0_0-\delta^2_0)=\frac{1-2|A_S|\sin\phi}{\eta^0_0}\,.
\end{eqnarray}

Since the interference between the $S$ and $D$ partial waves is
defined by the product $|A_S||A_D|\cos\delta$ and $\cos\delta$
determines $\delta$ only up to the sign, two solutions always
exist: the solution with $\delta>0$ and the other one with
$\delta<0$. Moreover, if $\cos\delta$ is close to 1 (and
$|\delta|\approx0$) in some region of $m$ then in this region a
transition from one solution to the other one is possible. The KEK
data [3] presented in Fig. 1c show that the phase $\delta$ changes
most rapidly near the $K\bar K$ threshold (which is one of the
evident manifestation of the $f_0(980)$ resonance) and that just
near 1 GeV $\cos\delta\approx1$. Thus, in principle, we have four
possible variants: (i) $\delta>0$ for $m<1$ GeV and $\delta<0$ for
$m>1$ GeV, (ii) $\delta>0$ for all $m$, (iii) $\delta<0$ for all
$m$, and (iv) $\delta<0$ for $m<1$ GeV and $\delta>0$ for $m>1$
GeV. However, variants (iii) and (iv) with $\delta<0$ for $m<1$
GeV can be rejected at once. Really, estimating $\delta$ for $m<1$
GeV by using the relation $\delta=\delta^0_0+\delta^2_0-\phi_D$,
one can easy verify that $\delta$ must be positive in this region
with the conventional definition of the signs of the phase shifts
$\delta^0_0$, $\delta^2_0$, and $\phi_D$ [see Figs. 1d, 2, and Eq.
(1)]. So, we shall consider only variants (i) and (ii).

Figure 3 shows the values of $\delta^0_0$ and $\eta^0_0$ extracted
from the KEK data (see Figs. 1a, 1b, and 1c) in the region
$0.68<m<1.64$ GeV by using the Eqs. (4) and (5) for the two above
mentioned variants of the $\delta$ phase behavior. The values of
$\delta^0_0$ in the region $0.36<m<1$ GeV obtained above from the
data on $|A_S|^2$ with $\eta^0_0=1$ (see Fig. 1d) are also shown
in Figs. 3a and 3b for comparison and completeness. As is seen,
for example, from Fig. 3a, the sets of the $\delta^0_0$ values
found in the region $0.68<m<1$ GeV by two different ways are in
quite reasonable agreement with each other. With obtaining
$\delta^0_0$ and $\eta^0_0$ in the general case, the Argand
diagram of the amplitude $\widetilde{A}_S$ was built for each
variant. After this the values of $2(\delta^0_0-\delta^2_0)$
obtained from Eqs. (4) and (5) were finally defined by the
requirement that those of $\delta^0_0$ be smoothly connected as a
function of $m$. That the strong violation of unitarity takes
place in variant (ii) for $m>1.16$ GeV (see Fig. 3d) can be easily
understood from the relation $\phi=\delta-2\delta^2_0+\phi_D$ [see
Eq. (4)]. The fact is that the values of $\phi$ in this case fall
into the range from $180^\circ$ to $360^\circ$, which is forbidden
as $\phi$ is the phase of the formally unitary amplitude
$\widetilde{A}_S$. Furthermore, in variant (ii), the phase sift
$\delta^0_0$ for $m>1$ GeV (see Fig. 3c) is in rather poor
agreement with the $\pi^+\pi^-$ production data [7-12] according
to which, for example, at $m\approx1.3$ GeV $\delta^0_0$ has to be
close to 270$^\circ$. Thus, variant (ii) with $\delta>0$ for all
$m$ can be rejected. As for variant (i), there are a set of
specific features which, to our knowledge, are missing from the
$\pi^+\pi^-$ data [7-12]. As is seen from Figs. 3a and 3b, in this
case we have noticeable differences of $\eta^0_0$ from unity for
$m<1$ GeV, its approximate equality to 1 for $1<m<1.12$ GeV,
violation of unitarity near 1.2 GeV, and sharp jumps of
$\delta^0_0$ and $\eta^0_0$ with further increasing $m$. There is
little doubt that these features are artefacts of the partial wave
analysis of the $\pi^-p\to\pi^0\pi^0n$ data.

Another difficulty is that the accepted normalization for
$|A_S|^2$ leads to $B_{f_2\pi\pi}=0.760\pm0.034$ [see Eq. (2)],
while according the PDG data [13]
$B_{f_2\pi\pi}=0.847\pm^{0.024}_{0.013}$. These two values cling
to one another only by their double errors. Hence, in principle,
one may conclude that the $\pi^0\pi^0$ data [3] indicate that the
absolute cross section of the $f_2(1270)$ resonance formation
through the OPE mechanism in the reaction $\pi^-p\to\pi^0\pi^0n$
can turn out to be approximately 20\% smaller, at least at
$m\approx m_{f_2}$, than that expected from the PDG data [13].
Alternatively, the KEK data [3] might be normalized with use of
the known value $\,\max|A_D|^2=(1+\eta^0_0)^2/4=B^2_{f_2\pi\pi}$
with $B_{f_2\pi\pi}$ from Ref. [13]. However, in this case, the
resulting values of $|A_S|^2$ in the most interesting region of
the lightest scalar resonance $\sigma(600)$ [3,13,19] would be
approximately 25\% higher than the unitarity limit for $|A_S|^2$.

We shall see in the next Sections that the other experimental data
on the reaction $\pi^-p\to\pi^0\pi^0n$ lead to very similar
difficulties.

\begin{center}{\bf III. ANALYSIS OF THE BNL DATA}\end{center}

In the BNL experiment [4], the high statistics on the reaction
$\pi^-p\to\pi ^0\pi^0n$ (about $8.5\times10^5$ events) has been
accumulated and the detailed partial wave analysis of the
$\pi^0\pi^0$ angular distributions has been performed. This
analysis has been done for ten sequential intervals in $-t$
covering the region $0<-t<1.5$ GeV$^2$ and over the $m$ range from
0.32 to 2.2 GeV scanned with the 0.04 GeV\,-wide step. As a
result, two solutions for the unnormalized intensities of the $S$
and $D_0$ partial waves and four ones (because of a sign
ambiguity) for their relative phase have been obtained. The above
quantities were denoted in Ref. [4] by $|S|^2$, $|D_0| ^2$, and
$\varphi_{S-D_0}$, respectively; in so doing, $D_0$ denotes the
$D$ wave with $L_z=0$, where $L_z$ is a projection of the
$\pi^0\pi^0$ relative orbital angular momentum on the $z$-axis in
the Gottfried-Jackson reference frame [4]. In the following, we
shall use these notations, too. One of the solutions for the $S$
and $D_0$ wave intensities, which is characterized by a large
magnitude of $|S|^2$ and a small one of $|D_0|^2$ for $m<1$ GeV,
and which is smoothly continued to the higher mass region, where
the $D_0$ wave is dominated by the $f_2(1270)$ resonance
contribution, has been selected in Ref. [4] as the physical
solution. Together with the intensities $|S|^2$ and $|D_0 |^2$,
the physical solution also includes two corresponding sets of the
$\varphi _{S-D_0}$ phase values which differ only in sign. Note
that the other solution intersects with the physical one at
$m\approx1$ GeV. We agree with the physical arguments given in
Ref. [4] based on which the other solution can be rejected in the
region $m<1$ GeV. However, for $m>1$ GeV we shell analyze the two
solutions and also the cases with transitions of the phase
$\varphi_{S-D_0}$ from one solution to the other one.

For the analysis we take the BNL data [4] on $|S|^2$, $|D_0|^2$,
and $\varphi_{S-D_0}$ pertaining to five intervals of $-t$,
$\,0.01<-t<0.03$ GeV$^2$, $0.03<-t<0.06$ GeV$^2$, $0.06<-t<0.1$
GeV$^2$, $0.1<-t<0.15$ GeV$^2$, and $0.15<-t<0.2$ GeV$^2$, and to
the region $0.32<m<1.6$ GeV. Note that the data on
$\varphi_{S-D_0}$ are available only for $m>0.8$ GeV. To obtain
the values of the quantities $|A_S|^2$, $|A_D|^2$, and $\delta$
(see Sec. II) as functions of $m$ characterizing the reaction
$\pi^+\pi^-\to\pi^0\pi^0$ on the mass shell, we parametrize the
$t$ dependence of $|S|^2$, $|D_0|^2$, and $\varphi_{S-D_0}$ by
means the following expressions
\begin{equation}
|S|^2=\frac{m^2}{q}|A_S|^2\frac{-t\exp[b_S(t-m^2_\pi)]}{(t-m^2_\pi)^2}\,,\
\ \ \
|D_0|^2=5\frac{m^2}{q}|A_D|^2\frac{-t\exp[b_{D_0}(t-m^2_\pi)]}{(t-m^2_\pi)^2}\,,
\end{equation} \begin{equation}
\varphi_{S-D_0}=\delta+\alpha\,(t/m^2_\pi-1)\,
\end{equation}
and, in each 0.04 GeV mass bin, thus determine the unnormalized
intensities $|A_S|^2$ and $|A_D|^2$, the phase $\delta$ and also
the slopes $b_S$, $b_{D_0}$, and $\alpha$ by fitting to the BNL
data on the $t$ and $m$ distributions by the formulae (6) and (7).
\footnote{Such two-parametric fits to the off-shall partial wave
intensities were widely used in the literature to obtain the
suitably extrapolated data (see, for example, Refs. [9,17,20,21].
However, the determination of the phase $\delta$ with use of the
direct extrapolation of the data on the phase $\varphi_{S-D_0}$
[see Eq. (7)] may provoke a question. If the data on the $S-D_0$
interference contribution, as such, had been presented in Ref.
[4], the problem would not have arisen. The fit to such data to
the function $-2ta\exp[b(t-m^2_\pi)]/(t-m^2_\pi)^2$ analogous to
those in Eq. (6) and the identification of the fitted parameter
$a$ with $\sqrt{5}(m^2/q)|A_S||A_D|\cos\delta$ would allow
$|\delta|$ to be determined in the proper way. Because such data
are not available, the indirect test of the results obtained with
Eq. (7) was carried out. Using the data [4] on $|S|^2$, $|D_0|^2$,
and $\varphi_{S-D_0}$ we constructed the quantity
$2|S||D_0|\cos\varphi_{S-D_0}$ and found with the above
extrapolation the on-shell $S-D$ interference contribution. Then,
knowing independently $|A_S|^2$ and $|A_D|^2$, we determined
$\delta$ as a function of $m$. The $\delta$ phase values obtained
in the two ways are in very close agreement with each other.
Certainly, owing to the forced double recounting of the errors of
the input data with the indirect test, the errors of $\delta$ turn
out to be larger then those obtained from the fit by Eq. (7). On
the other hand, when the values of $\delta$ are determined by
using Eq. (7) their errors practically are not differ from the
errors of the input data for $\varphi_{S-D_0}$. All the aforesaid
allowed us to prefer the determination of the phase $\delta$ with
use of Eq. (7).} In so doing, for $|S|^2$, $|D_0|^2$, and
$\varphi_{S-D_0}$ in each $-t$ bin we take into account the
physical solution  for $m<1$ GeV and the physical and other ones
for $m>1$ GeV. Unfortunately, the absolute value of the
$\pi^-p\to\pi^0\pi^0n$ reaction cross section has not been
determined in the BNL experiment [4]. Therefore, to normalize the
extrapolated intensities $|A_S|^2$ and $|A_D|^2$ we proceed in the
same way as in Sec. II. The extrapolated and normalized data
corresponding to the physical and other solutions are plotted in
Figs. 4a, 4c, and 4e with solid and open symbols, respectively. It
is interesting to note that as a result of the extrapolation two
branches of the $\varphi_{S-D_0}$ phase values for the other
solution (i.e., the branch with $\varphi_{S-D_0}>0$ for all $m$
and that with $\varphi_{S-D_0}<0$ for all $m$) interweave with
each other in the region $m>1.24$ GeV (see Fig. 4e) in such a way
that there arise two new branches of the extrapolated phase
$\delta$, which are characterized by a smooth dependence on $m$
and which, for example, can be considered either intersecting or
osculating near 1.26 GeV.

As in Sec. II, we begin with the determination of the phase shift
$\delta^0_0$ for $m<1$ GeV from the data on $|A_S|^2$ (see Fig.
4a) assuming that $\eta^0_0=1$ in this region. The resulting phase
shift values are shown in Fig. 5 by open circles. Note that two
points in the region $m\approx m_K$ disturbed by the
$K^0_S\to\pi^0\pi^0$ events [4] are omitted. Then, heaving the
data on $|A_S|^2$, $|A_D|^2$, and $\delta$ for $m>0.8$ GeV (see
Fig. 4), we determine the values of $\delta^0_0$ and $\eta^0_0$
with use of the general formulae (3) and (4), and also Eq. (1).
The results for the previously selected solutions among all the
possible ones, that are shown in Fig. 4, are plotted in Fig. 5
with solid circles. Strictly speaking, the selection is reduced to
rejection of the physical solution with $\delta<0$ for $m<1$ GeV
(see Fig. 4e), since a simple estimate
$\delta^0_0=\delta-\delta^2_0+\phi_D$ (see Sec. II) yields
$\delta^0_0\approx-(25\div40)^\circ$ for this solution in the
region of $0.8-1$ GeV, which is, certainly, unsatisfactory. In its
turn, Figs. 5a and 5b show that the physical solution with
$\delta>0$ for all $m$ can be also rejected due to strong
violation of the unitarity condition for $m>1.2$ GeV. Figures 5c
and 5d correspond to the physical solution for $|A_S|^2$,
$|A_D|^2$, and $\delta$ with the transition of the phase $\delta$
at $m\approx1$ GeV from the branch with $\delta>0$ to that with
$\delta<0$ (see Fig. 4e). Such a physical solution consists with
unitarity but corresponds to the weak coupling between the
$\pi\pi$ and $K\bar K$ channels near the $K\bar K$ threshold.
Indeed, for this solution $\eta^0_0$ is close to unity in the
region $1<m<1.15$ GeV. However, the latter disagrees with the data
obtained from the reactions $\pi^-p\to\pi^+\pi^-n$,
$\pi^+p\to\pi^+\pi^-\Delta^{++}$, and $\pi N\to K\bar K(N,\Delta)$
(see, for example, Refs. [7,9,10,16,22,23]). Figures 5e and 5f
correspond to the combination of the physical solution with
$\delta>0$ for $m<1$ GeV and the other one for $m>1$ GeV with
$\delta>0$ in the region $1<m<1.28$ and $\delta<0$ in the region
$1.28<m<1.52$ GeV (see Figs. 4a and 4e). Finally, Figs. 5g and 5h
correspond to the similar combination of the physical solution and
the other one for which $\delta<0$ for $m$ from 1 to 1.52 GeV (see
also Fig. 4e). Certainly, there are two more variants which differ
from the last two ones only by the sign of $\delta$ for $m>1.28$
GeV (see Fig. 4e). These variants lead, however, to appreciable
violations of the unitarity condition for $m>1.32$ GeV, and
therefore, are of little interest. Thus, one can conclude that
just the variant presented in Figs. 5e and 5f is, in many
respects, in qualitative agreement with the results of the
previous partial wave analyses of the $\pi^+\pi^-$ data
[1,7,9,11]. As indicated above, this variant corresponds to the
positive relative phase $\delta=\phi_S-\phi_D$ up to the
$f_2(1270)$ resonance and the negative one above it. An important
point is that such a behavior of $\delta$ as a function of $m$ is
strongly confirmed by the pioneering data from the polarized
target experiment on the reaction $\pi^-p\to\pi^+\pi^-n$ at 17.2
GeV [11].

Comparing Fig. 5 with Fig. 3 we just note that the BNL data lead
to the obviously higher values of the phase shift $\delta^0_0$ for
$m<0.5$ GeV than the KEK data.

In extracting the information on $\delta^0_0$ and $\eta^0_0$, the
phase $\phi_D$ has been defined by fitting to the data on
$|A_D|^2$ (see Fig. 4c) with use of Eq. (1). The parameters of the
$f_2(1270)$ were found to be (see also the curves on Fig. 4c): for
the physical solution,
\begin{eqnarray}m_{f_2}=1.279\pm0.002\ \mbox{GeV}\,,& \qquad & \Gamma_{f_2}
=0.205\pm0.005\ \mbox{GeV}\,,\nonumber\\ R_{f_2}=3.96\pm0.24\
\mbox{GeV}^{-1}\,,& \qquad & B_{f_2\pi\pi}=0.697\pm0.008
\end{eqnarray} and, for the other solution,
\begin{eqnarray}m_{f_2}=1.281\pm0.002\ \mbox{GeV}\,,& \qquad & \Gamma_{f_2}
=0.211\pm0.005\ \mbox{GeV}\,,\nonumber\\ R_{f_2}=4.65\pm0.33\
\mbox{GeV}^{-1}\,,& \qquad & B_{f_2\pi\pi}=0.712\pm0.007\,.
\end{eqnarray} Thus, the BNL data indicate
that the branching ratio $B_{f_2\pi\pi}$ can amount to
approximately 84\% of the PDG value [13]. Possible consequences of
a similar discrepancy has been already discussed in connection
with the KEK data at the end of Sec. II (recall that the relevant
ratio for the KEK data has been found to be approximately 90\%).

\begin{center}{\bf IV. DISCUSSION OF THE GAMS DATA}\end{center}

The highest statistics on the reaction $\pi^-p\to\pi^0\pi^0n$ was
accumulated by the GAMS Collaboration in two experiments at 38 GeV
[5] and 100 GeV [6]. However, the small $-t$ region from 0 to 0.2
GeV$^2$ has been examined in the works [5,6] very sparingly. But
the data averaged over the small $-t$ region have been presented
for $|S|^2$ and $\varphi_{S-D_0}$ in Ref. [5] and for $|S|^2$,
$|D_0|^2$, and $\varphi_{S-D_0}$ in Ref. [6]. Such ``global"\
data, of course, do not permit to perform a proper extrapolation
of the mass distributions measured from the physical region to the
pion pole. Nevertheless, we discuss some typical features of the
GAMS data. The physical solution for $|S|^2$ and $\varphi_{S-D_0}$
and the other solution only for $|S|^2$ have been presented in
Ref. [5] in the region $0.8<m<1.6$ GeV. The $\varphi_{S-D_0}$
phase was found to be positive in the full mass range [5] (about
existing the ambiguous solution with $\varphi_{S-D_0}<0$ the
readers, probably, have to guess by themselves). In general, the
available GAMS data [5] are very similar to the corresponding BNL
data [4]. For example, in the case of the physical solution,
$|S|^2$ and $\varphi_{S-D_0} $ from Ref. [5] behave as functions
of $m$ in the same way as the extrapolated quantities $|A_S|^2$
and $\delta$ shown in Figs. 4a and 4e by solid circles. However,
it is such a physical solution for $|A_S|^2$ and $\delta$ (with
$\delta>0$ for all $m$) that leads to strong violation of the
unitarity condition for $m>1.2$ GeV (see Fig. 5b). In analyzing
the $\pi^0\pi^0$ system produced in the reaction
$\pi^-p\to\pi^0\pi^0n$ at 100 GeV, the only solution for $|S|^2$,
$|D_0|^2$, and $\varphi_{S-D_0}$ has been selected and presented
by the GAMS Collaboration in Ref. [6]. Unfortunately, this unique
solution is very close to the above physical one obtained in the
GAMS 38 GeV $\pi^-p\to\pi^0\pi^0n$ experiment [5].

It is of first importance that the GAMS Collaboration measured the
absolute cross section of the $f_2(1270)$ resonance formation in
the $D_0$ wave in the region $0<-t<0.2$ GeV$^2$. According to Ref.
[24], at 38 GeV, $\sigma_{D_0}(\pi^-p\to
f_2(1270)n\to\pi^0\pi^0n)=2.3\pm0.2$ $\mu$b. More recently, this
value was used, in particular, to normalized the 100 GeV data [6].
Although the cross section value obtained is approximately 1.5--2
times greater than in a set of previous $\pi^0\pi^0$ production
experiments [24,25], nevertheless, it is 1.57 times smaller than
the estimate based on the OPE model (it is an old story with the
experimental underestimation of the $\pi^-p\to
f_2(1270)n\to\pi^0\pi^0n$ reaction cross section all details of
which can be found in Ref. [25]). By using this model with the PDG
values of $m_{f_2}$, $\Gamma_{f_2}$, and $B_{f_2\pi\pi}$ [13] we
estimate
$$\sigma_{D_0}(\pi^-p\to
f_2(1270)n\to\pi^0\pi^0n)\approx\sigma^{OPE}(\pi^-p\to
f_2(1270)n\to\pi^0\pi^0n)\approx$$ \begin{equation} \approx\
\frac{g^2_{\pi^-pn}}{4\pi}\,\frac{5\pi}{m_p^2P^2_{\pi^-}}\,m_{f_2}
\,\Gamma_{f_2}\,\frac{2}{9}\,B^{\,2}_{f_2\pi\pi}\int\limits_{-0.2\mbox{{\small
GeV}}^2}^{0} \frac{-t\exp[b_{f_2}(t-m^2_\pi)]}{(t-m^2_\pi)^2}\
dt\,\approx\,3.6\,\mu\mbox{b}\,,\end{equation} where
$P_{\pi^-}=38$ GeV, $g^2_{\pi^-pn}/4\pi\approx2\times14.3$, and
$b_{f_2}\approx7.5$\,GeV$^{-2}+2\times0.8$\,GeV$^{-2}\ln(38/
18.3)$ $\approx8.68$\,GeV$^{-2}$ (in estimating the slope
$b_{f_2}$, its Regge energy dependence and the results for the
slope $b_{D_0}$ in the $f_2(1270)$ mass region presented in Fig.
4d have been taken into account). Note that this estimate is in
good agreement with the result of the extrapolation of the
available data on the reaction $\pi^-p\to
f_2(1270)n\to\pi^+\pi^-n$ at 17.2 GeV [8], 100 GeV and 175 GeV
[26] to the GAMS energy (see Ref. [25] for details). Thus, the
GAMS data [24] indicate that the value of $B_{f_2\pi\pi}$ can
amount to about 80\% of that given by the PDG [13].

We now summarize briefly the common difficulties encountered in
analyzing the data obtained in four recent experiments on the
reaction $\pi^-p\to\pi^0\pi^0n$. First, the physical solutions
selected by using the partial wave analyses of the $\pi^0\pi^0$
production data lead to the values of $\delta^0_0$ and $\eta^0_0$
which are incompatible with the known results obtained from the
$\pi^+\pi^-$ data, at least for $m>1$ GeV. Some of these solution
lead to strong violations of the unitarity condition. On the other
hand, among the other solutions one can point out, in principle,
the more preferred ones. Secondly, it is astonishing that the data
of the four recent experiments on $\pi^0\pi^0$ production include
the indications that the value of $B_{f_2\pi\pi}$ can be
distinctly smaller than the currently accepted one. This
difficulty is rather serious and highly interesting. Let us remind
that the $\pi\pi$ production experiments on unpolarized targets,
in particular, those under discussion here, do not permit the
contributions of the $\pi$ and $a_1$ exchange mechanisms to be
separated in principle, even with huge statistics, because these
contributions to the unpolarized cross section are incoherent
[27]; in other words, there is no model-independent way to do
this. Therefore, by our opinion, the difficulty with
$B_{f_2\pi\pi}$ may present itself a further evidence that the
partial wave analyses of the unpolarized data allow to determine
the intensities and the relative phases of the $S$, $D$, ...
$\pi\pi$ partial waves only approximately, with any extrapolation
method. ``The degree of proximity"\ is associated with the
relative magnitude of the non-leading $a_1$ exchange contribution.
With high statistical accuracy of the unpolarized data the
presence of the $a_1$ exchange mechanism can manifest itself in
the events responsible for $|S|^2$ just in the form of the above
difficulty. In fact, this statement follows naturally from the
analysis of the unnormalized KEK and BNL data (see the end of Sec.
II and also Sec. V for details). As for the GAMS data [23], they
appear to point merely to the general problem involving the
accurate measurement of the $\pi^-p\to\pi^0\pi^0n$ reaction cross
section. More discussions both of the additional assumptions
needed in analyzing the unpolarized data and of the $a_1$ exchange
contribution can be found in the works [1,2,11,12,27-29].

\begin{center}{\bf V. CONCLUSION}\end{center}

Using the most simple way we have extracted the values of the
$I=0$ $\pi\pi$ $S$ wave phase shift $\delta^0_0$ and inelasticity
$\eta^0_0$ from the current data on the reaction
$\pi^-p\to\pi^0\pi^0n$. It seems clear that a new set of precise
experiments on this reaction is needed both for the more precise
definition of the $\pi^0\pi^0$ production mechanism and for
obtaining the more detailed information on $\pi\pi$ scattering and
light scalar resonances in the $\pi\pi$ channel. Let us formulate
in this connection several concrete suggestions, leaving aside the
general wish to investigate the reaction $\pi^-p\to\pi^0\pi^0n$ on
the polarized target.

1) Detailed data on the $m$ and $t$ distributions for the
$\pi^0\pi^0$ $S$ and $D_0$ partial waves, especially in the region
$0<-t<0.2$ GeV$^2$ where the OPE mechanism dominates, and
measurements of the absolute value of the $\pi^-p\to\pi^0\pi^0n$
reaction cross section at different energies, for example, at KEK,
BNL, IHEP, and CERN, would be highly desirable. The relative
accuracy of new measurements must be comparable with (or better
of) that given by the PDG [13] for $B^2_{f_2\pi\pi}$. This would
allow to perform the accurate description of the $f_2(1270)$
formation differential cross section within the OPE model and to
test how well the $S$ wave $\pi^0\pi^0$ production cross section
at its absolute maximum (which is located in the region
$0.6<m<0.8$ GeV) agrees with the OPE model prediction under the
standard normalization condition according to which $|A_S|^2= 1$
(i.e., $\delta^0_0-\delta^2_0=90^\circ$ and $\eta^0_0=\eta^2_0=1$)
at the absolute maximum point. An excess of the experimental
values over the model expectations would be a good evidence,
obtained from the unpolarized target data, for the presence of the
$a_1$ exchange contribution to the $S$ wave $\pi^0\pi^0$
production cross section in the region of its absolute maximum.
Alternatively, if the maximal experimental value of the $S$ wave
cross section turns out to be less than in the OPE model, then it
will completely disturb of the existing ideas about the
$\delta^0_0$ phase shift for $m<1$ GeV, which seems to be highly
unlikely.

2) We suggest to perform in the low $-t$ region the especially
careful measurements of $\pi^0\pi^0$ production in the $S$ wave
for $m$ from 0.9 to 1.1 GeV, i.e., in the region of the well known
interference minimum in $|S|^2$ located near the $K\bar K$
threshold. This would allow to obtain the important additional
information on the $f_0(980)$ resonance coupling constant to the
$K\bar K$ channel, $g_{f_0K\bar K}$, and to resolve the
long-standing question [16] concerning a possible ambiguity in the
behavior of the phase shift $\delta^0_0$ above the $K\bar K$
threshold which arises at $g^2_{f_0K^+K^-}/4\pi>4\pi
m^2_K\approx3.1$ GeV$^2$. Furthermore, the magnitude of the $S$
wave intensity in the immediate region of the minimum (if it lies
below the $K\bar K$ threshold) can be used to obtain a very strong
upper limit on the $a_1$ exchange contribution at small $-t$ in
this region of $m$.

3) As a rule, the assumption of phase coherence between the $D_0$
and $D_-$ amplitudes is one of those using to select the physical
solution, see, for example, Refs. [4,6,9,30]. Here $D_-$ denotes
the $D$ wave with $|L_z|=1$, in the Gottfried-Jackson reference
frame, which is produced via the unnatural parity exchanges in the
$t$ channel of the reaction $\pi N\to\pi\pi N$. In this connection
we would like to call attention to a new curious circumstance.
According to the GAMS measurements [6], the ratio
$|D_-|^2/|D_0|^2$ in the $f_2(1270)$ mass region at 100 GeV is
half as large as that at 38 GeV. This fact may testify to
compensation of the $\pi P$ and $a_2 P$ Regge cut contributions
($P$ denotes the Pomeron exchange) to the $D_-$ wave production
amplitude with energy increasing, i.e., to violation of phase
coherence.

4) We also suggest to use in future the more suitable notation for
the $S-D_0$ interference contribution to be extracted in the
unpolarized target experiments, instead of the commonly used
simplified one of the form $|S||D_0|\cos\varphi_{S-D_0}$. It
includes the mention of the coherence factor (see, for example,
Ref. [31]) and is more adequate to the measured quantity.
Experimentally, the $S$ and $D_0$ wave intensities, $|S|^2$ and
$|D_0|^2$, and the $S-D_0$ interference contribution,
$\,\xi\,|S||D_0|\cos\widetilde{\varphi}$, are measured
simultaneously. In fact, $|S|\equiv[|S_\pi|^2+|S_{a_1}|^2]^{1/2}$,
$\,|D_0|\equiv[|D_{0\pi}|^2+|D_{0a_1}|^2]^{1/2}$, and the
coherence factor $\xi\,$ $(0\leq\xi\leq1)$ and the phase
$\widetilde{\varphi}\,$ have the form
$$\xi=|\sum_{i=\pi,a_1}S_iD^*_{0i}|/[(\sum_{i=\pi,a_1}|S_i|^2)
(\sum_{i=\pi,a_1}|D_{0i}|^2)]^{1/2}\,,$$
$$\widetilde{\varphi}=\arctan[(\sum_{i=\pi,a_1}|S_i||D_{0i}|\sin\varphi_i)/
(\sum_{i=\pi,a_1}|S_i||D_{0i}|\cos\varphi_i)]\,,$$ where
$S_\pi\,(D_{0\pi})$ and $S_{a_1}\,(D_{0a_1})$ are the $S\,(D_0)$
wave production amplitudes caused by the $\pi$ and $a_1$ exchange
mechanisms, respectively (these amplitudes correspond to two
independent configurations of the nucleon helicities in the
reaction $\pi N\to\pi\pi N$), and $\varphi_i$ is the relative
phase between the amplitudes $S_i$ and $D_{0i}$. Let us consider
the case when the amplitude $D_{0a_1}$ is negligible. Then,
denoting $\widetilde{\varphi}=\varphi_\pi$ by $\varphi_{S-D_0}$,
one can see that the real interference contribution differs from
that presented with the simplified notation,
$|S||D_0|\cos\varphi_{S-D_0}$, by the coherence factor
$\xi=1/\sqrt{1+|S_{a_1}|^2/|S_\pi|^2}$. If we put $\xi=1$ for all
$m$, we shall always deal with the effectively underestimated
values of $|\cos\varphi_{S-D_0}|$.

\begin{center}{\bf ACKNOWLEDGMENTS}\end{center}

We would like to thank the E852 Collaboration for allowing free
access to the detailed BNL data, http://dustbunny,
physics.indiana.edu/pi0pi0pwa/. This work was supported in part by
the grant INTAS-RFBR No. IR-97-232 and the grant RFBR No.
02-02-16061.


\newpage\begin{figure}\centerline{\epsfysize=7.2in\epsfbox{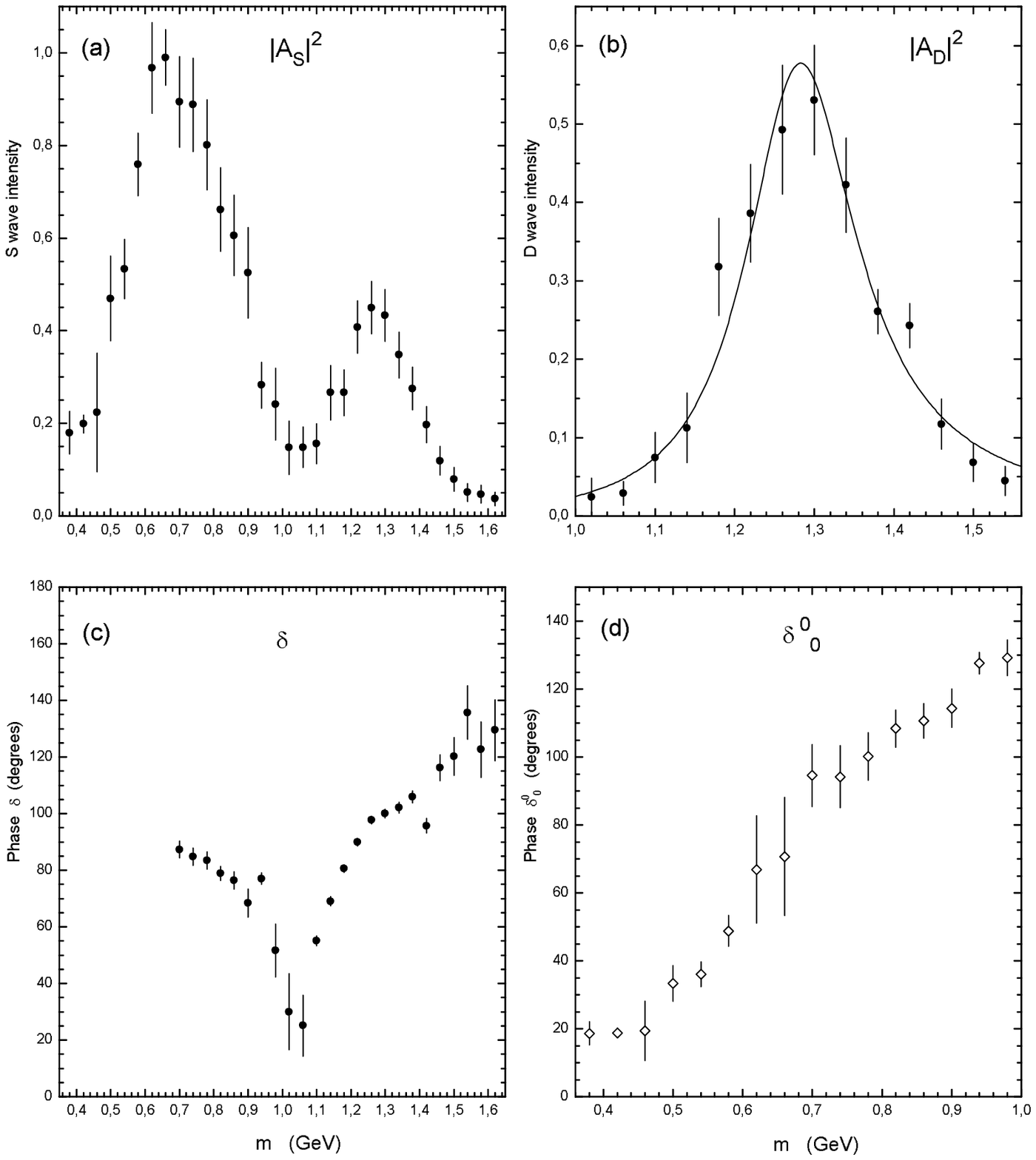}}
\vspace*{0.3cm} \caption{\footnotesize{(a), (b), (c) The KEK data
on the reaction $\pi^+\pi^-\to\pi^0\pi^0$ [3]. (a) The normalized
$S$ wave intensity $|A_S|^2$. (b) The normalized $D$ wave
intensity $|A_D|^2$; the curve is the fit using Eq. (1) with the
parameters of the $f_2(1270)$ presented in Eq. (2). (c) The
relative phase $\delta$ between the amplitudes $A_S$ and $A_D$.
(d) The $I=0$ $\pi\pi$ $S$ wave phase shift $\delta^0_0$ obtained
from the data on $|A_S|^2$ alone under the assumption $\eta^0_0$
is unity.}}\end{figure}

\newpage\begin{figure}\centerline{
\epsfxsize=6in\epsfbox{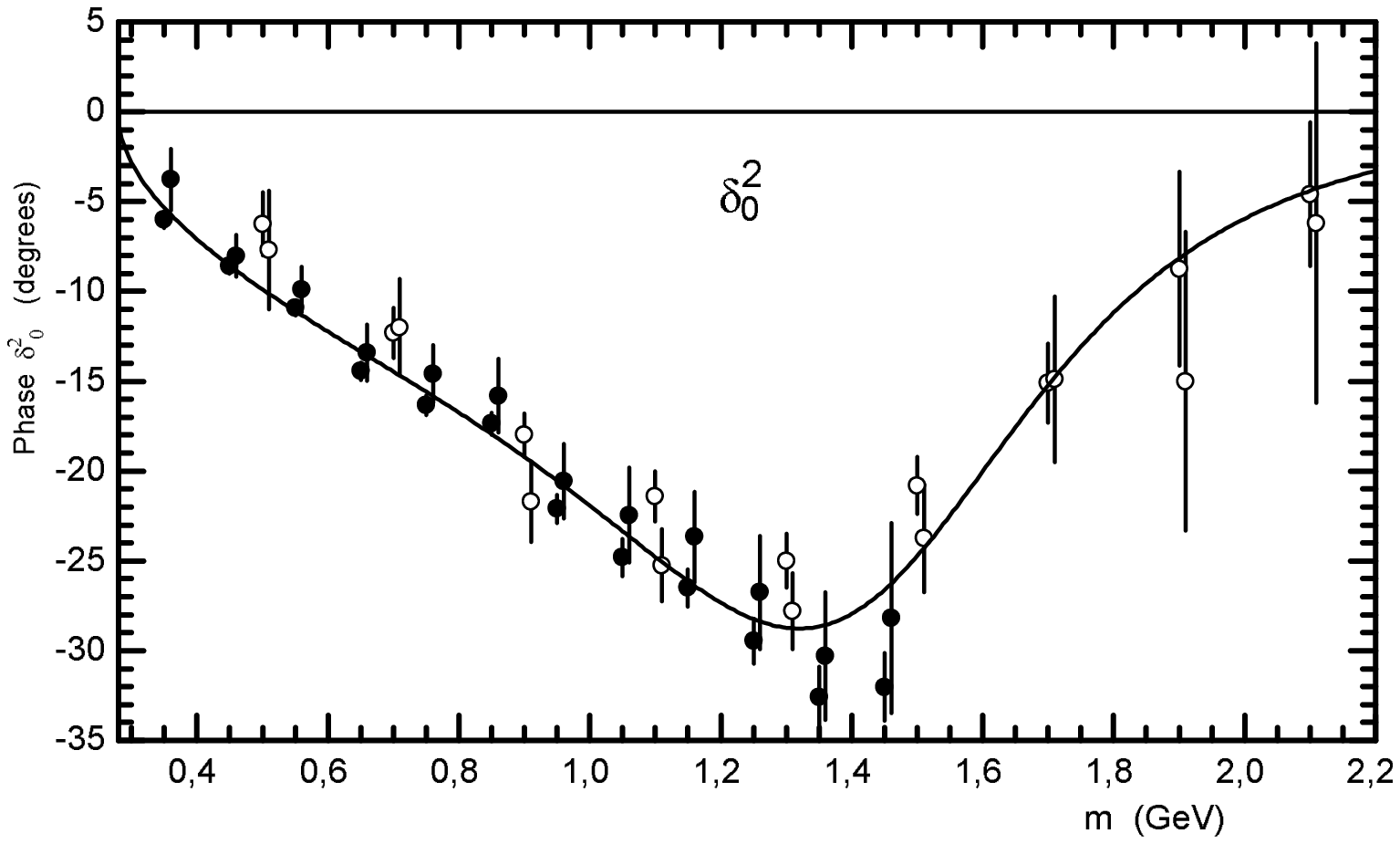}}
\vspace*{0.3cm}\caption{\footnotesize{The $I=2$ $\pi\pi$ $S$ wave
phase shift $\delta^2_0$. The data are from Refs. [17] (solid
circles) and [18] (open circles). The curve corresponds to the fit
described in the text.}}\end{figure}

\newpage\begin{figure}\centerline{\epsfysize=7.2in\epsfbox{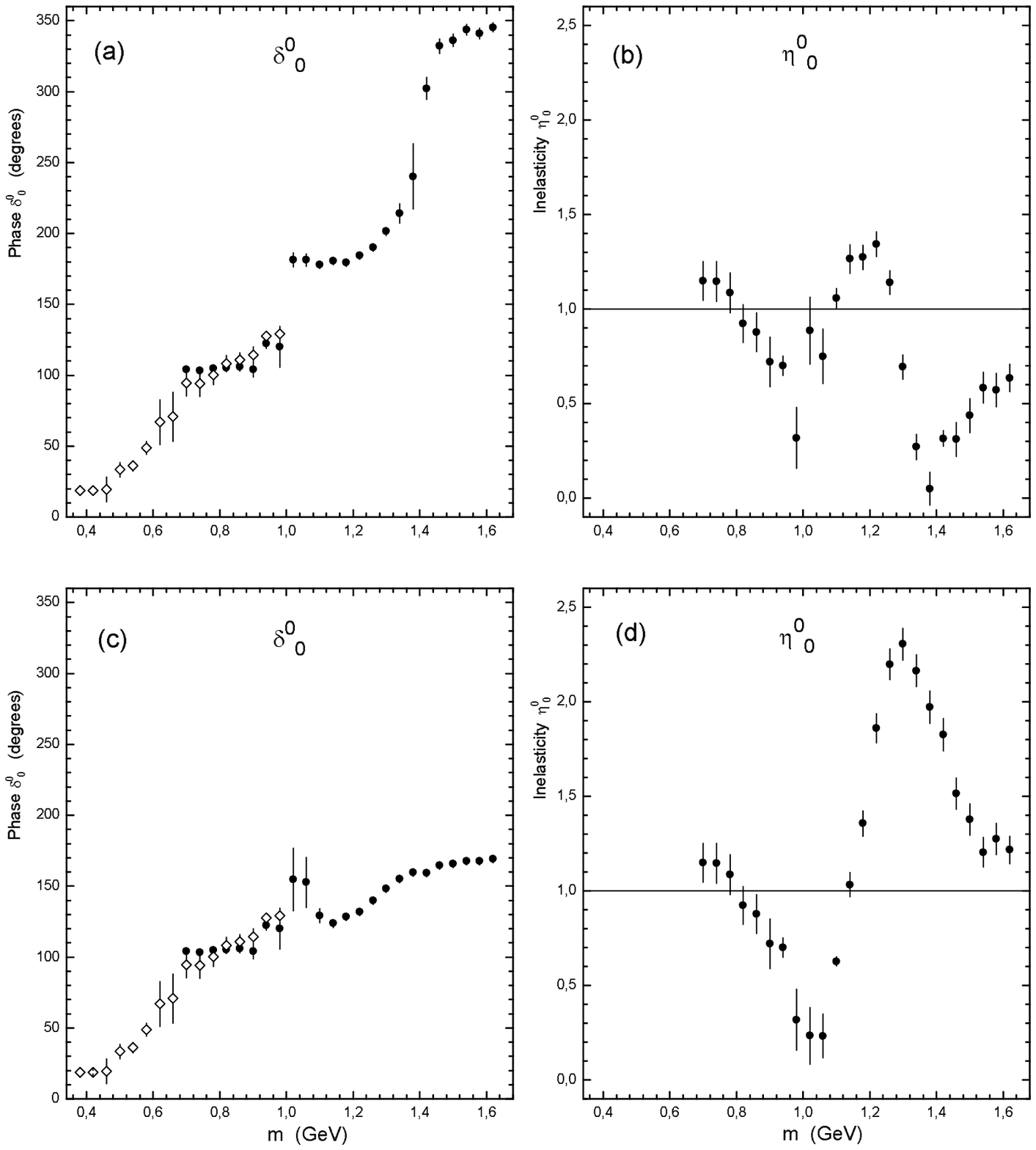}}
\vspace*{0.3cm}\caption{\footnotesize{The solid circles show the
values of the phase sift $\delta^0_0$ (a) and inelasticity
$\eta^0_0$ (b) extracted from the KEK data [3] on the reaction
$\pi^+\pi^-\to\pi^0\pi^0$ in the case that the $m$ dependence of
the phase $\delta$ corresponds to variant (i) described in the
text. (c), (d) The same for variant (ii). The open diamonds show
the values of $\delta^0_0$ corresponding to Fig. 1d.}}\end{figure}

\newpage\begin{figure}\centerline{\epsfysize=7.2in\epsfbox{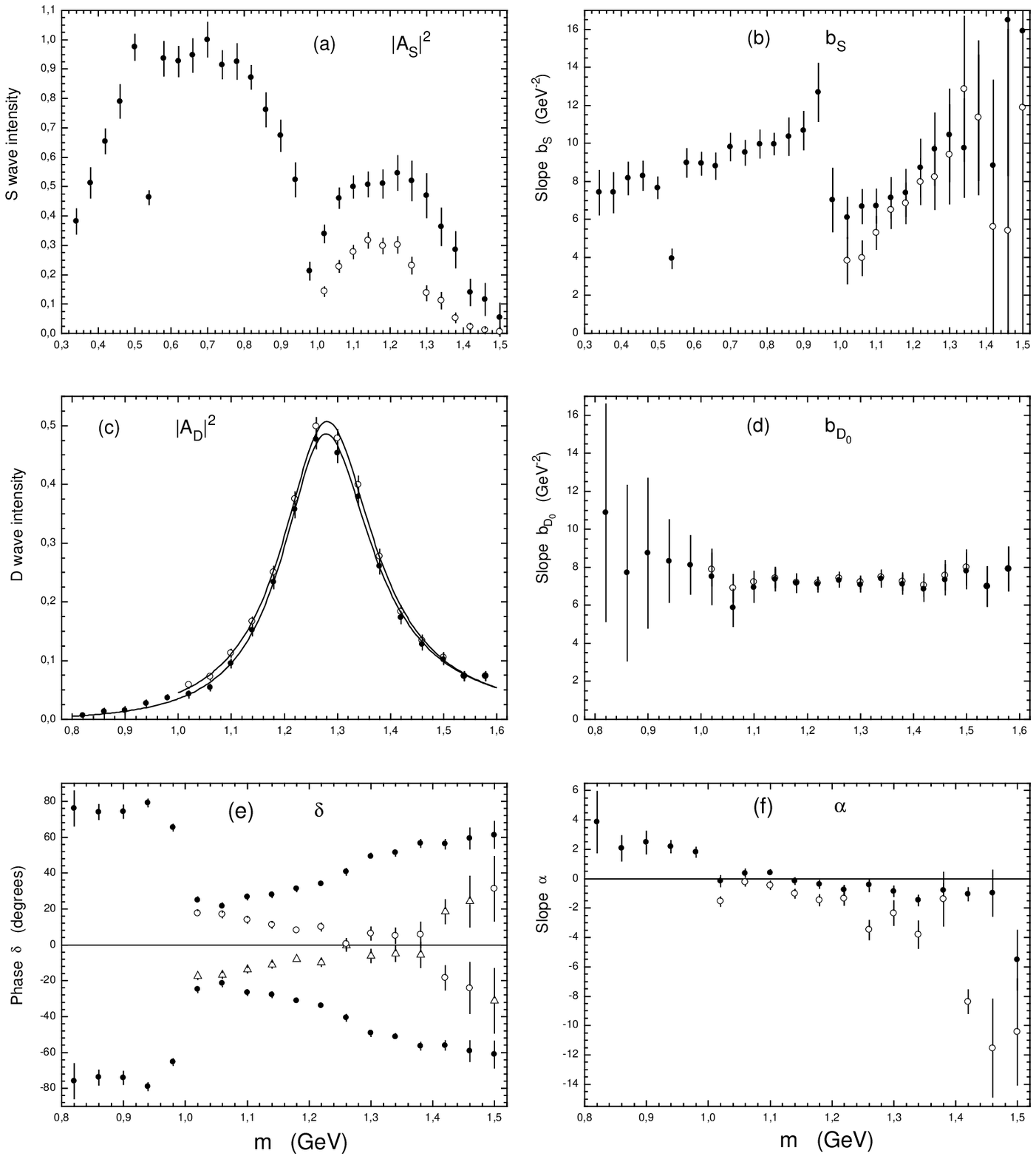}}
\vspace*{0.3cm}\caption{\footnotesize{The results of the
extrapolation of the BNL data [4]. (a) The extrapolated and
normalized $S$ wave intensity. (c) The extrapolated and normalized
$D$ wave intensity. (e) The extrapolated relative phase $\delta$
between the $S$ and $D$ wave amplitudes. The slopes $b_S$ (b),
$b_{D_0}$ (d), and $\alpha$ (f) as functions of $m$. The solid
circles correspond to the physical solution. The open circles [and
also the open triangles in plot (e) for $\delta$] correspond to
the other solution. The lower and upper curves in plot (c) are the
fits using Eq. (1) with the parameters of the $f_2(1270)$
presented in Eqs. (8) and (9), respectively. }}\end{figure}

\newpage\begin{figure}\centerline{\epsfysize=7.2in\epsfbox{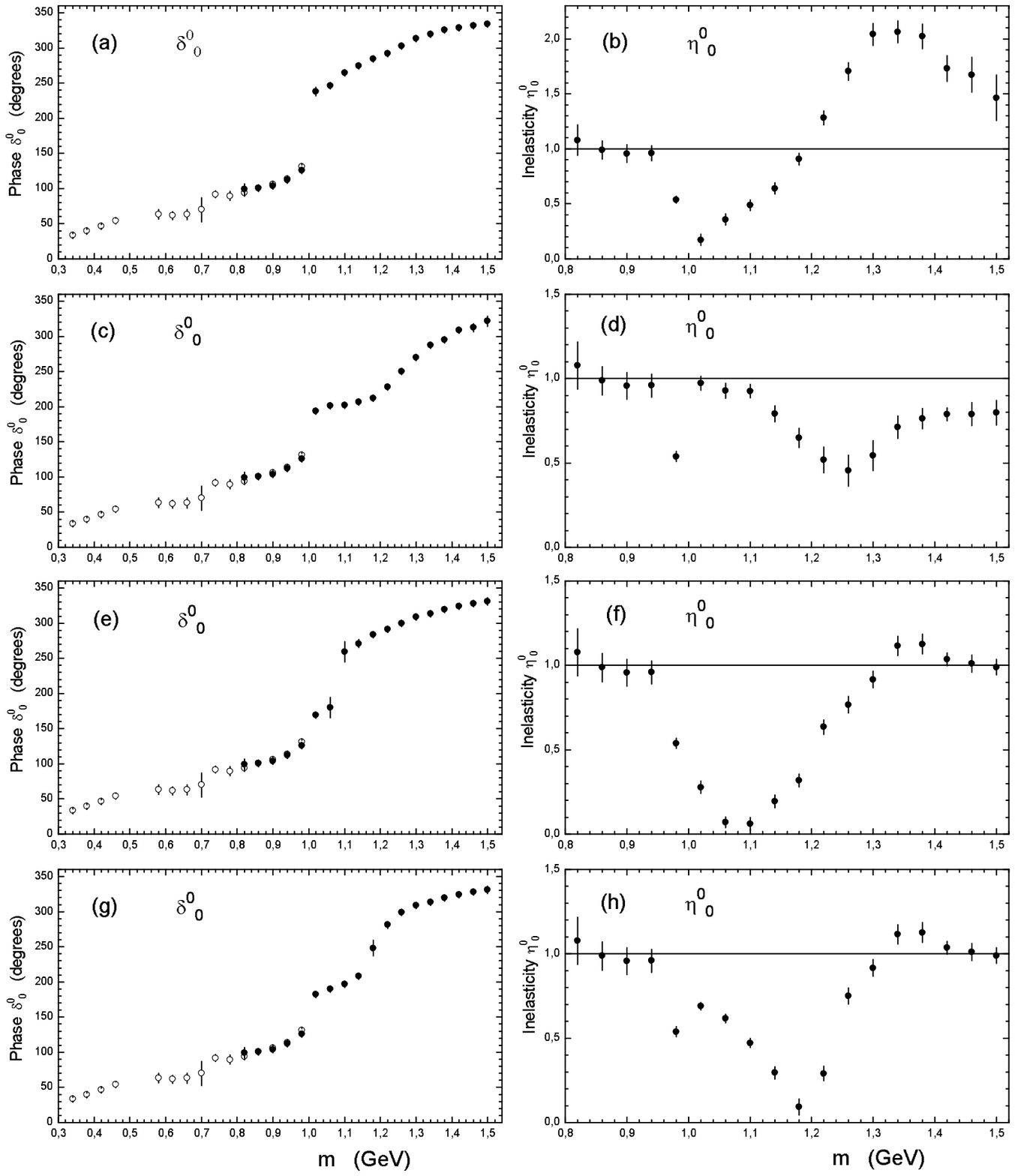}}
\vspace*{0.3cm}\caption{\footnotesize{The phase shift $\delta^0_0$
and inelasticity $\eta^0_0$ extracted from the BNL data [4]. Plots
(a) and (b) correspond to the physical solution (see Fig. 4) with
$\delta>0$ for all $m$. Plots (c) and (d) correspond to the
physical solution with a transition of the phase $\delta$ at
$m\approx1$ GeV from the branch pertaining to its positive values
to that with its negative ones (see Fig. 4e). Plots (e) and (f)
correspond to the combination of the physical solution with
$\delta>0$ for $m<1$ GeV and the other solution for $m>1$ GeV with
$\delta>0$ in the region $1<m<1.28$ GeV and with $\delta<0$ in the
region $1.28<m<1.52$ GeV (see Figs. 4a and 4e). Plots (g) and (h)
correspond to the combination of the physical solution with
$\delta>0$ for $m<1$ GeV and the other solution for $m>1$ GeV with
$\delta<0$ in the region $1<m<1.52$ GeV (see Figs. 4a and 4e). The
open circles show the values of the phase shift $\delta^0_0$
obtained from the data on $|A_S|^2$ for $m<1$ GeV (see Fig. 4a)
alone under the assumption $\eta^0_0$ is unity.}}\end{figure}

\end{document}